\documentclass[10pt,letterpaper]{article}
\usepackage[top=0.85in,left=2.75in,footskip=0.75in,marginparwidth=2in]{geometry}

\usepackage{amsmath}
\usepackage{amsthm}
\usepackage{graphicx}

\usepackage{cite}

\usepackage{microtype}
\DisableLigatures[f]{encoding = *, family = * }

\raggedright
\usepackage{changepage}

\usepackage[aboveskip=1pt,labelfont=bf,labelsep=period,singlelinecheck=off]{caption}

\makeatletter
\renewcommand{\@biblabel}[1]{\quad#1.}
\makeatother

\begin{document}
\vspace*{0.35in}

\begin{flushleft}
{\Large
\textbf\newline{Kinetic theory and Brazilian income distribution}
}
\newline
\\
Igor D. S. Siciliani\textsuperscript{},
Marcelo H. R. Tragtenberg\textsuperscript{1*},
\\
\bigskip
\bf{1} Universidade Federal de Santa Catarina-UFSC
\\
\bigskip
* marcelo.tragtenberg@ufsc.br

\end{flushleft}

\section*{Abstract}
We investigate the Brazilian personal income distribution using data
from National Household Sample Survey (PNAD), an annual research available
by the Brazilian Institute of Geography and Statistics (IBGE). It
provides general characteristics of the country's population. Using
PNAD data background we also confirm the effectiveness of a semi-empirical
model that reconciles Pareto power-law for high-income people and
Boltzmann- Gibbs distribution for the rest of population. We use three
measures of income inequality: the Pareto index, the average income
and the crossover income. In order to cope with many dimensions of
the income inequality, we calculate these three indices and also the
Gini coefficient for the general population as well as for two kinds
of population dichotomies: black/ indigenous /mixed race versus white
/ yellow; and men versus women. We also followed the time series of
these indices for the period 2001-2014. The results suggest a decreasing
of Brazilian income inequality over the selected period. Another important
result is that historically-disadvantaged subgroups (Women and black
/ indigenous / mixed race),that are the majority of the population,
have a more equalitarian income distribution. These groups have also
a smaller monthly income than the others and this social structure
remained virtually unchanged in the period of time.

\section{A simple model of income distribution}

The first power-law has discovered about a century ago by economist
Vilfredo Pareto long before the personal income distribution analysis
to become an important meaning in the econophysics. Pareto stated
there is a simple law which governs all distribution of income (at
least for the asymptotic high-income regions)\cite{s2006}: 

\begin{equation}
P(m)=\frac{A}{m^{(1+\alpha)}}
\end{equation}
then, applying $\log$ on both sides of the expression, $\log(P(m))=\log(A)-(1+\alpha)\log(m)$,
where A is a constant , $m$ represents the personal income and $\alpha$
is known as Pareto Index. It means that log-log plotted of Pareto
power-law obey a linear behavior\cite{ms1997}. As a consequence,
the higher is the value of $\alpha$, the lower will be the income
inequality between the wealthiest people.

After the Pareto power-law discovery it has been exhaustively confirmed
by researches conducted in different countries, types of society at
various periods of history over the following years.

Other extremely useful model to study of income distribution and wealth
inequality in a determinate population of agents with an arbitrary
individual income $m_{i}$ is possible by analogy with the Boltzmann-Gibbs
distribution (BGD). In classical kinetic theory the BGD is the most
probable distribution $\overline{f(\varepsilon)}$ for an equilibrium
gas of $N$ elements enclosed in a box. This consideration is independent
of particle shape or details about interactions that occurs during
elastic collisions (energy conservation), in this sense BGD is universal. 

The transposition of this physical knowledge to the economy world
happens when energy $\varepsilon$ is replaced by money and for this
reason the cash is also conserved. Therefore an initial arbitrary
money distribution function must meet the conditions: 

\begin{equation}
\sum_{i=1}^{N}\phi_{i}=\varPhi
\end{equation}
\begin{equation}
\sum_{i=1}^{N}m_{i}\phi_{i}=M
\end{equation}
where we defined $\phi_{i}$ as the number of agents and $M$ as total
income.

In order to satisfy that two specific needs, the stationary distribution
of income $P(m)$ should be exponential as equilibrium BGD. Mathematically
after a few steps: 

\begin{equation}
P(m)=C\exp(-\lambda m_{i})
\end{equation}
where $C$ is a constant and $\lambda$ is a Lagrange multiplier.
In this way, normalizing $P(m)$ to unity: $\int_{0}^{\infty}P(m)dm=1$
and then calculating the average income $<m>=\int_{0}^{\infty}mP(m)dm$.
Finally, we can rewrite the distribution function of personal income
as,

\begin{equation}
P(m)=\frac{1}{T}\exp(-m/T)
\end{equation}
note that the ``temperature'' T is equal to the average income $\left\langle m\right\rangle $\cite{Dr_gulescu_2001}. 

An overview of some empirical studies of wealth distribution indicates
that the exponential or log-normal behavior are observed for 90-95\%
of the population and Pareto power-law for the wealthiest rest of
population\cite{Clementi_2005,s2006}.

On the basis of this information we propose a general stationary income
distribution function for a population:

\begin{equation}
P(m)=\frac{1}{T}\theta(m_{l}-m)\exp(-m/T)+\frac{A\theta(m-m_{l})}{m^{(1+\alpha)}}
\end{equation}
in the equation above $\theta$ is the Heaviside function, $m_{l}$
is the crossover income of personal income which the BGD is observed
and $A$ is a experimental constant. The effectiveness of that model
became clear from the situations discussed in this text. Next for
convenience, from equation $\left(6\right)$ we can derive the cumulative
probability distribution $P(m'\geq m):$

\begin{equation}
P(m'\geq m)=\int_{m}^{\infty}P(m)dm
\end{equation}
hence, $P(m'\geq m)=\exp(-m/T)-\exp(-m_{l}/T)+\frac{Am_{l}^{-\alpha}}{\alpha}$
if $m\leq m_{l}$ and $P(m'\geq m)=\frac{Am^{-\alpha}}{\alpha}$ if
$m\geq m_{l}$ . Recognizing that in the vast majority of cases $\alpha\sim1.5$
and which obviously $\frac{m_{l}}{T}>1$, we estimate $\left(7\right)$
as

\begin{equation}
P(m'\geq m)\cong\begin{cases}
\begin{array}{c}
\exp(-m/T)\\
Bm^{-\alpha}
\end{array} & \begin{array}{c}
m\leq m_{l}\\
m\geq m_{l}
\end{array}\end{cases}
\end{equation}
to wit, the probability of a person having an individual income equal
or greater than to $m$. Thus, we can recognize that equation $(8)$
is very closely linked to the econophysics \textquoteleft \textquoteleft two-class\textquoteright \textquoteright{}
theory of Yakovenko\cite{S2014}.

\section{Income inequality measures}

Through previous semi-empirical model we can take three useful parameters
to measure income inequality between subgroups of a same population
or even to compare different populations, the Pareto index, the ``temperature''
and the crossover income $m_{l}$. The Pareto index to be directly
associated with wealthiest people and so it is limited by this range.
But $\alpha$ is still very convenient to make comparisons and even
analyzing its temporal evolution as can be seen in some studies\cite{Silva_2005,s2002,cg2005}. 

The ``temperature'' and the crossover income can provide us a good
perspective about income inequality between subgroups belonging to
a given population at a same time; however, due to exchange rates
between the base currency and foreign currencies, inflation and other
complex aspects the average income and crossover income without adequate
treatment can not contribute much more than that.

Beyond the proposed model the Gini coefficient (G) has been the most
popular instrument used to measure income inequality in the literature\cite{wikigini,De_Maio_2007}. 

\begin{figure}[h!]
Lorenz curve

\noindent\begin{minipage}[t][1\totalheight][c]{1\columnwidth}%
\includegraphics[width=75mm]{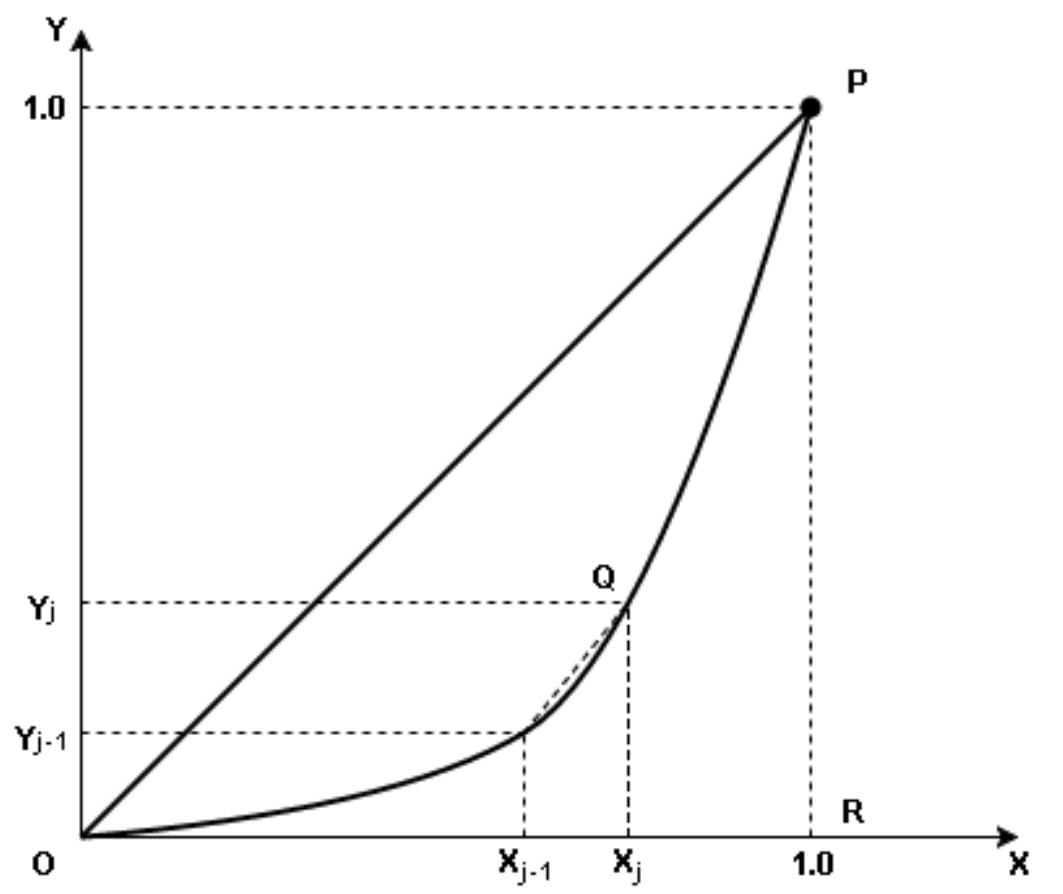}%
\end{minipage}

\noindent\begin{minipage}[t][1\totalheight][c]{1\columnwidth}%
{\footnotesize{}Figure 1: Lorenz curve shown in fractional scale:
X is cumulative percentage of population and Y is the cumulative percentage
of total income.}%
\end{minipage}
\end{figure}

In Figure 1 the simple function $Y=X$ represents the situation where
the income is perfectly distributed, i.e., the ``poorest'' 20\%
of population would earn the same percentage of total income and so
on\cite{De_Maio_2007}. The other function commonly known as Lorenz
curve shows the observed behavior in real societies. So, the Gini
coefficient can be obtained through calculating the ratio of the area
between the diagonal line and Lorenz curve (OPQ) by area of the triangle
beneath diagonal (OPR). In other words, $G=\frac{AreaOPQ}{AreaOPR}$.
As a result, the key problem becomes estimating the value of the area
below the Lorenz curve. Numerically this result can be achieved by
successive approximation of trapezoids $S_{j}$. Note that for N pairs
(X,Y) under the Lorenz curve we can build N-1 trapezoids\cite{hd1972}.
The jth trapezoid base is $Y_{j}$ and $Y_{j-1}$whose height is $(X_{j}-X_{j-1})$.
Thus, 
\begin{equation}
S_{j}=\frac{(Y_{j}+Y_{j-1})(X_{j}-X_{j-1})}{2}
\end{equation}
in view of previous equation we can define graphically the Gini coefficient
by Brown's formula:

\begin{equation}
G=1-2\sum_{J}^{N-1}S_{j}=1-\sum_{J}^{N-1}(Y_{j}+Y_{j-1})(X_{j}-X_{j-1})
\end{equation}
while there are many positive contributions to the Gini coefficient
in the evaluation of income inequality, it also has a few disadvantages.
The most evident disadvantage is concerned with its highly sensitive
to transfer especially of the middle classes\cite{w2010}.

\section{Income distribution in Brazil }	

In this section, we investigate the Brazilian personal income distribution
using microdata from National Household Sample Survey (PNAD) available
by the Brazilian Institute of Geography and Statistics (IBGE). The
PNAD is an annual research that study general aspects of Brazilian
society, regarding labor, income, education and others\cite{ibge}.
Under the circumstances, we extracted from PNAD data the value of
monthly income from all sources for people aged over 10 years$(m)$
and also for four subsets: black / indigenous / mixed race, white
/ yellow, man and woman. After we got the income variable $m$ and
we neglected the persons without income as well as missing values
in the new subset and once we begin making the analysis and fitting
functions to the distributions. In this way, a part of the results
of the paper is about economically active population.

In the specific case of the year 2014 we had 362,625 cases of which
following the criteria mentioned earlier we worked with 219,288 of
them. Initially to provide evidences of the effectiveness of model
described here we calculated the cumulative probability distribution
of income for all Brazilian population as you can see on the Figure
2.

\begin{figure}[h!]
PNAD of 2014

\noindent\begin{minipage}[c]{1\columnwidth}%
\includegraphics[width=100mm]{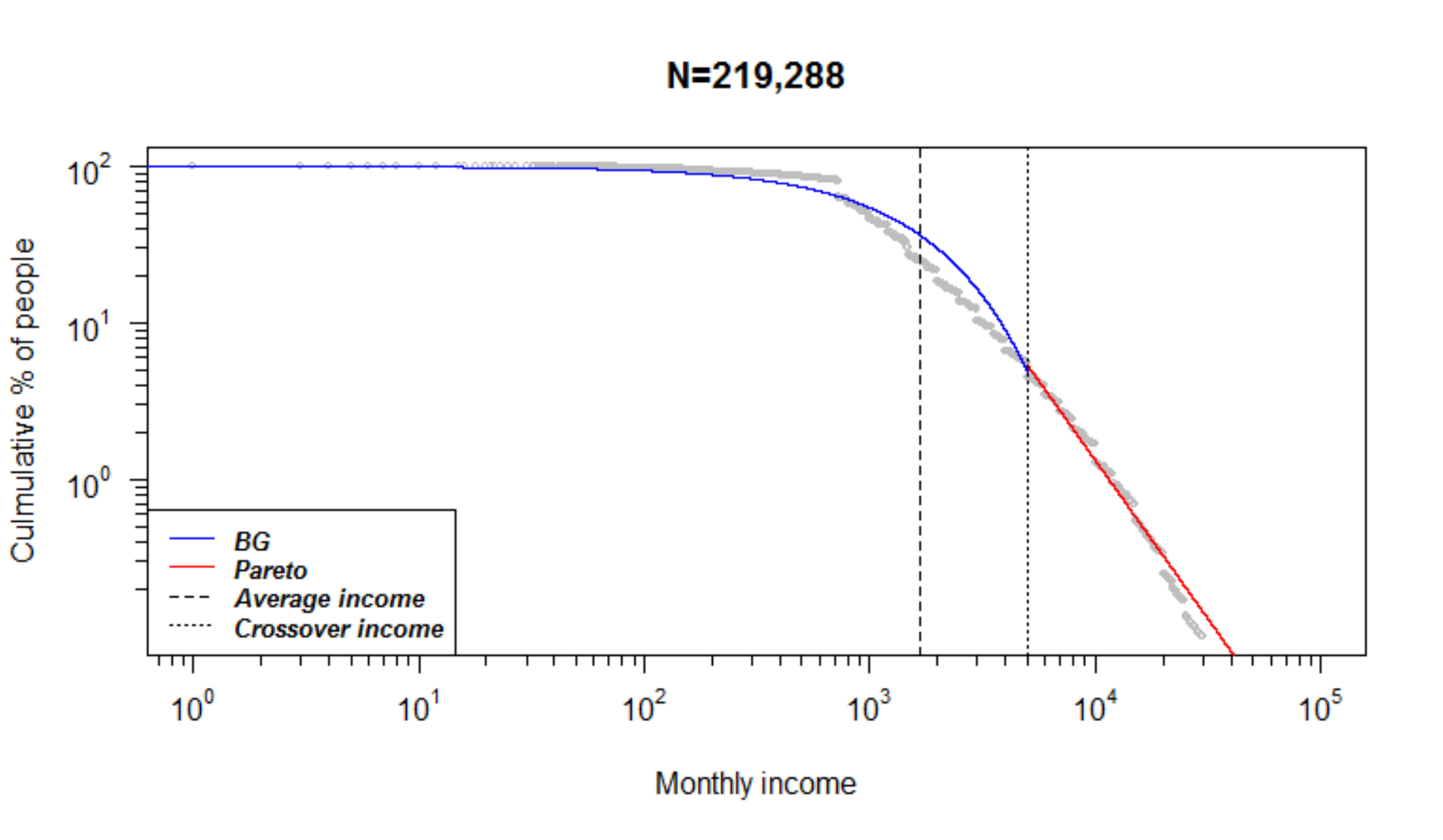}%
\end{minipage}

\noindent\begin{minipage}[t][1\totalheight][c]{1\columnwidth}%
{\footnotesize{}Figure 2: Cumulative probability distribution of income
on a log-log scale. The grey points represents the empirical values
extracted from 2014 PNAD and the solid lines corresponds to model
fit (equation 8), where power-law region is differentiate on a red
color which it represents almost 5 \% of population. The graph rest
obeys Boltzmann-Gibbs distribution (blue line) . As regards the parameters
to measure income inequality, the average income is highlighted by
the dashed vertical line with value is $1655\pm6R\$$ and the dotted
vertical line is the crossover income $5000R\$$. We also obtained
$\alpha=2.181\pm0.011$ and $m_{l}=5000R\$$ for the other coefficients
calculated.}%
\end{minipage}
\end{figure}

The Pareto index of $2.181$, its standard error of $0.011$ and also
a 95\% confidence interval of $0.024$ around this empirical measured
were estimated following bootstrap procedure on half number of randomly
selected cases in the wealthiest region (5\% of total population),to
this end we resampling 40 times the initial randomly sample.

To complete this inequality parameters set of the Brazilian population
we calculate the Gini coefficient associating the Brown's formula
and the method of convergent extrapolation oscillation\cite{F2008}.
This methodology allowed us to verify that the measure of the degree
of uncertainty associated with this estimate is virtually nil due
to extreme convergence. For this reason, we just estimate the accuracy
of measure in the fist divergent decimal case. Hence, $G=0.504$ for
total population through 2014 PNAD data.

\begin{figure}[h!]
Gini coefficient from 2014 PNAD

\begin{minipage}[c]{0.45\columnwidth}%
\includegraphics[width=75mm]{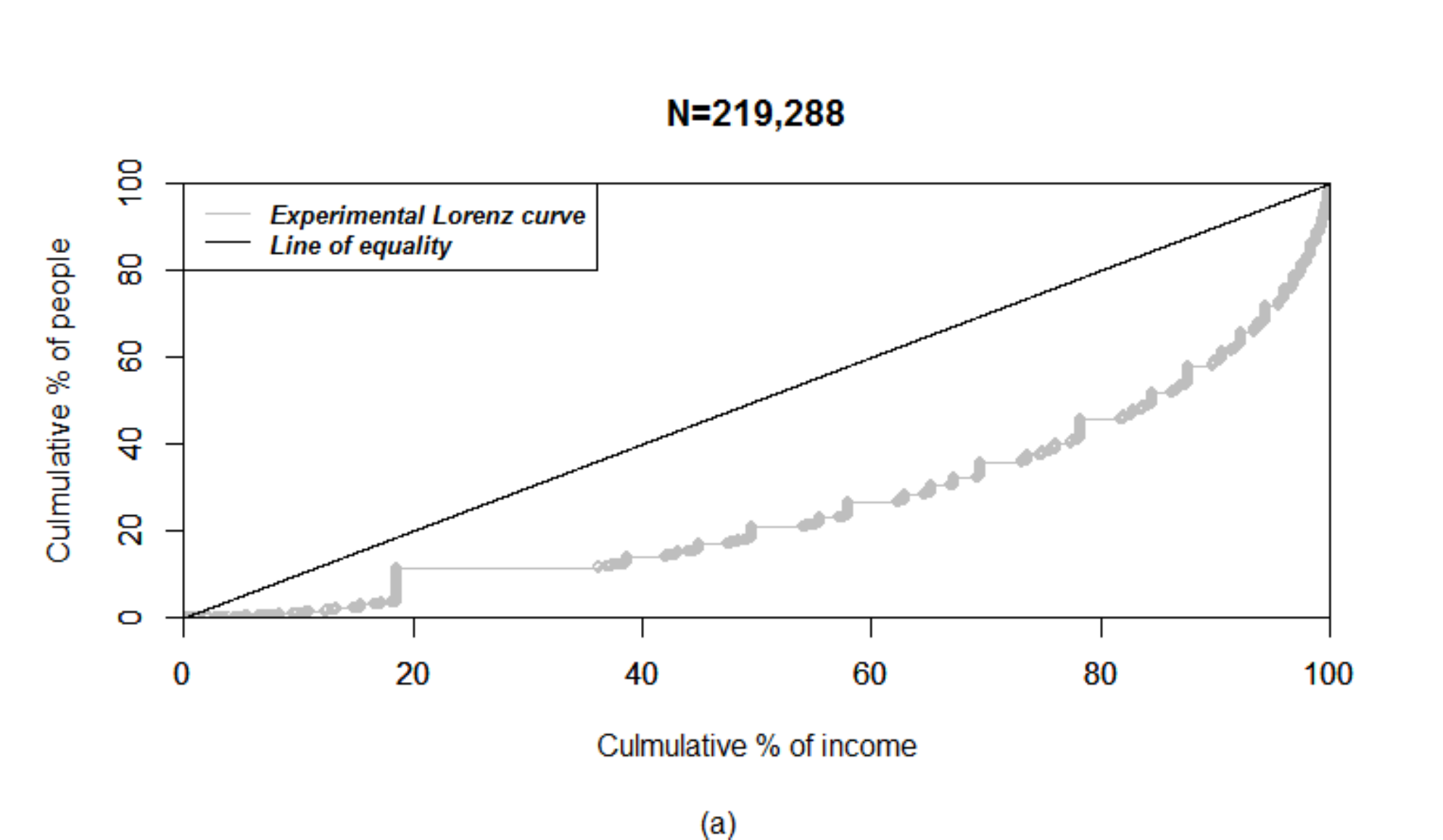}%
\end{minipage}
\begin{minipage}[c]{0.45\columnwidth}%
\includegraphics[width=50mm]{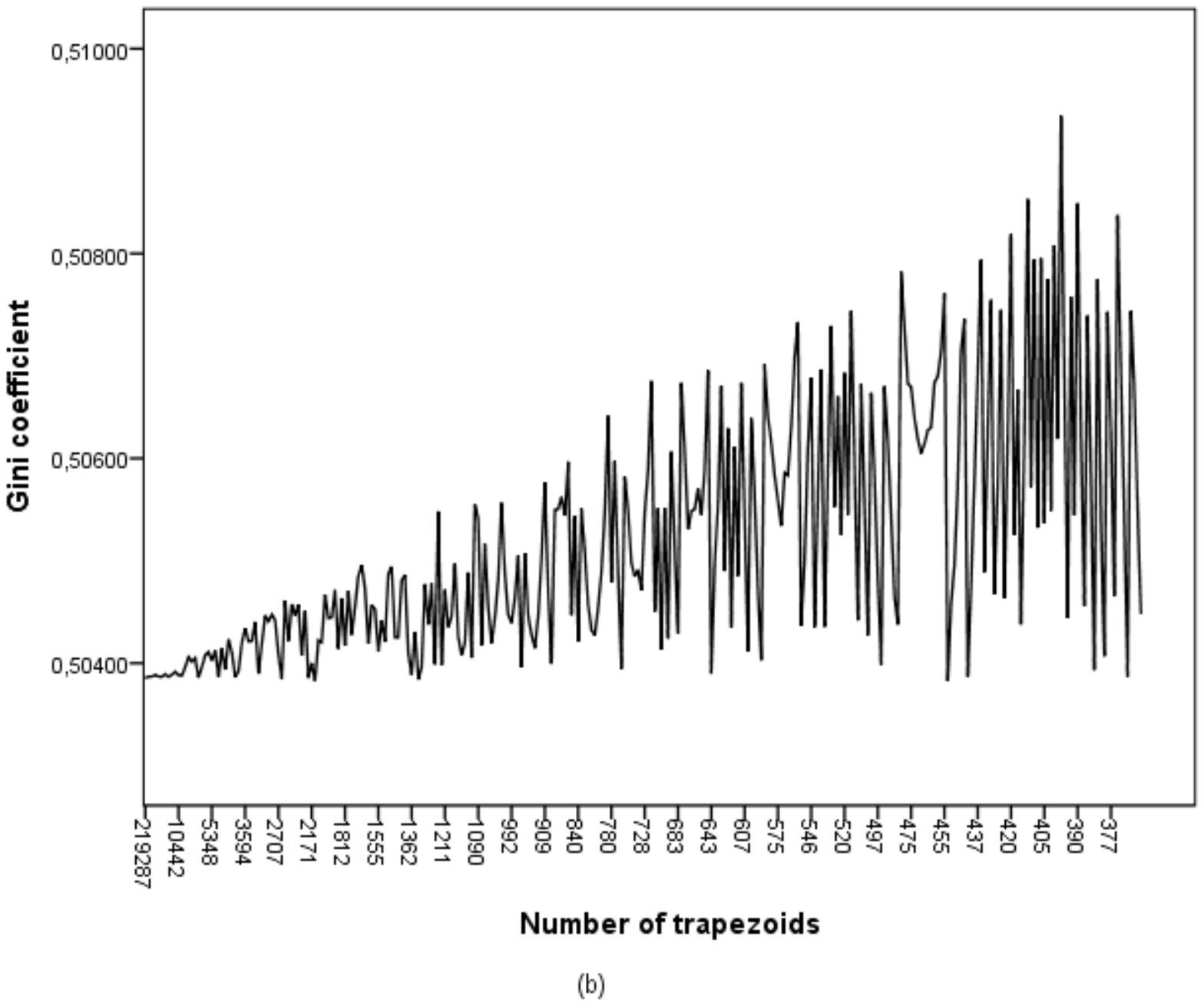}%
\end{minipage}
\smallskip{}
\noindent\begin{minipage}[c]{1\columnwidth}%
{\footnotesize{}Figure 3: (a) Empirical Brazilian population Lorenz
curve. (b) Method of convergent extrapolation oscillation.}%
\end{minipage}
\end{figure}

We made the same analysis for two different dichotomies of the population.
The first is the division by gender (man and woman) and the second
is the division by color/race (black / indigenous / mixed race and
white / yellow). Then, we compare the inequality parameters for each
of this cases. 

\begin{table}[!h]
\noindent\begin{minipage}[t][1\totalheight][c]{1\columnwidth}%
\begin{tabular}{lllll}
\multicolumn{5}{c}{PNAD of 2014}\tabularnewline
\hline 
Group & \multicolumn{2}{l}{Sex/Gender} & \multicolumn{2}{l}{Race/Color}\tabularnewline
Subgroup & Man & Woman & WY & BIM\tabularnewline
Gini coefficient & 0.497 & 0.497 & 0.519 & 0.457\tabularnewline
Temperature (R\$) & $1940\pm9$ & $1359\pm6$ & $2144\pm11$ & $1278\pm5$\tabularnewline
Pareto index & $2.172$$\pm0.012$ & $2.247\pm0.020$ & $2.306\pm0.017$ & $2.187\pm0.012$\tabularnewline
crossover income (R\$) & 6000 & 4010 & 7000 & 3510\tabularnewline
\end{tabular}%
\end{minipage}

\noindent\begin{minipage}[c]{1\columnwidth}%
{\footnotesize{}Table 1: Inequality parameters for each subgroup of
2014 PNAD, where BIM means black, indigenous, mixed race and WY means
white, yellow. To construct the table we clean data using the same
criteria applied to total population. Likewise, we fixed with good
precision the }crossover income{\footnotesize{} ($m_{l}$) as lower
limit to income of the 5\% wealthiest population of each subgroup
like we made for total population.}%
\end{minipage}
\end{table}

As can be seen from Table 1, for the gender group there is no difference
between the Gini coefficient measured to men and women. For the 5\%
richest people we see a small improvement in economically active woman's
income distribution if we compare it with the man's income distribution.
This modest improvement is statistical significantly if we consider
the 95\% confidence interval of 0.024 around both man's and woman's
Pareto index. However, the differences between men and women are more
expressively if we focus on the monthly income values, since both
the crossover income and the ``temperature'' are much higher for
men. This contrast between the income parameters of men and women
can be better understood if we note that in the year of 2014 through
the PNAD data, about 51.5\% of the respondents were women, but only
40\% of the total monthly income belonged to them. When we analysis
the population separation by color, we note that the inequality in
general is lower for the BIM subgroup. But joining to this result
the differences between BIM and WY temperatures that it represents
almost 68\% of the BIM average monthly income, and the crossover income
contrast at the same subgroups, we can recognize that the lower Gini
coefficient and higher Pareto index calculated to BIM subgroup reflects
the fact that the people belonging to this subset are concentrate
at the poorer classes be either BG region or Pareto region. For the
group separated by color 57.5\% of the PNAD respondents are self-identified
as black, indigenous or mixed-race while about 44\% of the total monthly
income is earned by BIM members.

\section{Time series of inequality coefficients}

To broaden our understanding of Brazilian income distribution, in
this section we investigate the dynamics of that inequality parameters
over the years 2001-2014. We use the same PNAD data source as before
and still we maintain the criteria of data cleaning, statistical procedure
and formation of groups previously adopted for the 2014 PNAD alone.

The first interesting observation is about the Pareto region that
remained fairly constant over the selected years ranging from 4\%
to 6\% of total population. For this reason, we fixed the crossover
income as the lower monthly income of the 5\% richest people that
is represents asymptotic high-income region. The next result relates
to the time series of the values obtained for temperature and crossover
income of the total population, both measured in local currency (Real). 

\begin{figure}[h!]
Time series of temperature and crossover income

\noindent\begin{minipage}[c]{1\columnwidth}%
\includegraphics[width=100mm]{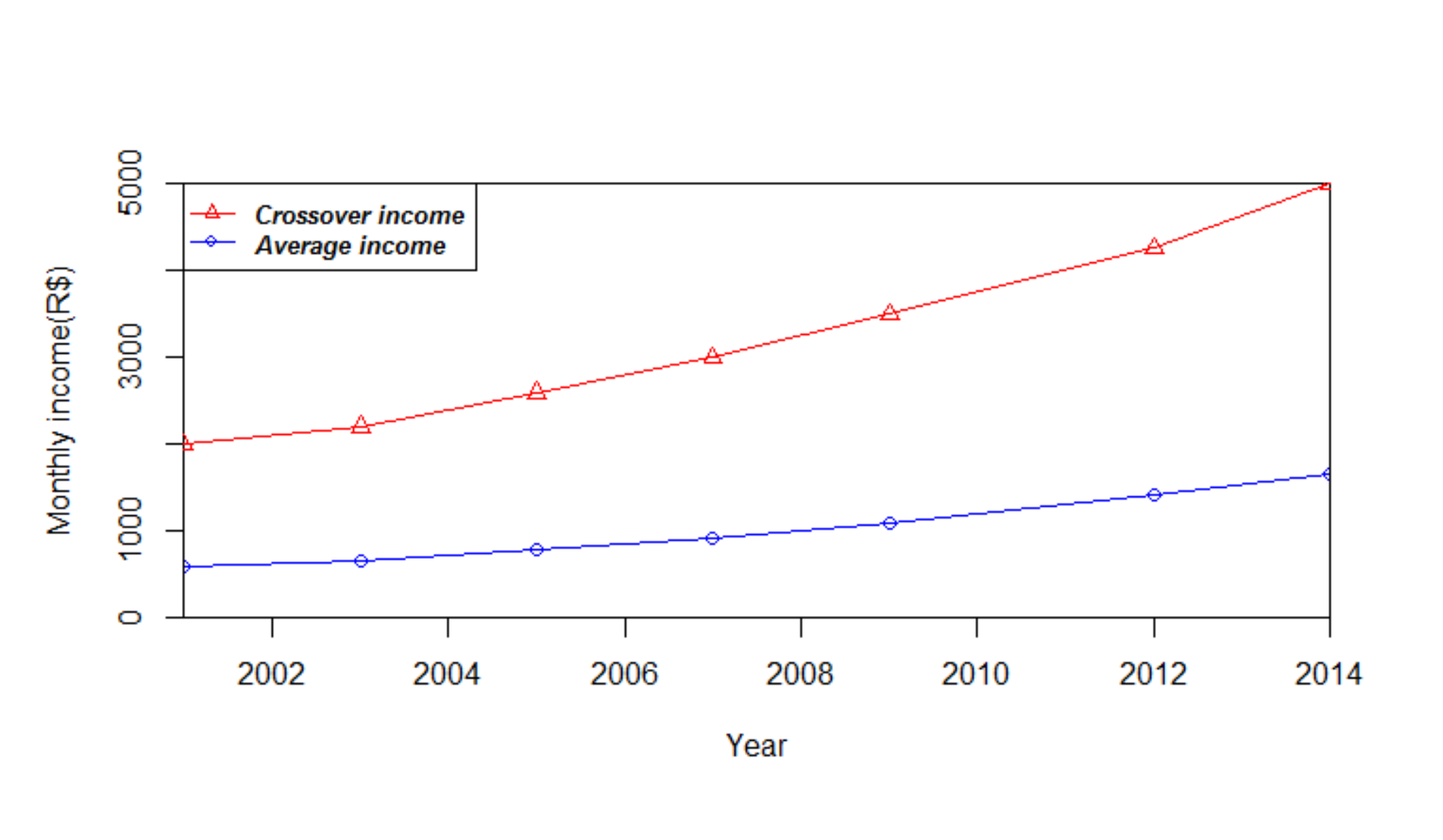}%
\end{minipage}

\noindent\begin{minipage}[c]{1\columnwidth}%
{\footnotesize{}Figure 4: Times series of the average and }crossover
income{\footnotesize{} calculated for total population through the
year 2001, 2003, 2005, 2007, 2009, 2012, 2014.}%
\end{minipage}
\end{figure}
In this context, if we analyze the Figure 4 we can observe that the
percentage difference between the ``temperature'' and the crossover
income in the year of 2001 corresponded to 241\% of the average income.
For the last year studied (2014) this percentage difference decreases
to 202\%.

Focusing on the time series of Gini coefficient, we estimated it for
the same subsets as before: gender (man, woman) and color (BIM, WY). 

\begin{figure}[h!]
Time Series of Gini coefficient

\noindent\begin{minipage}[c]{1\columnwidth}%
\includegraphics[width=100mm]{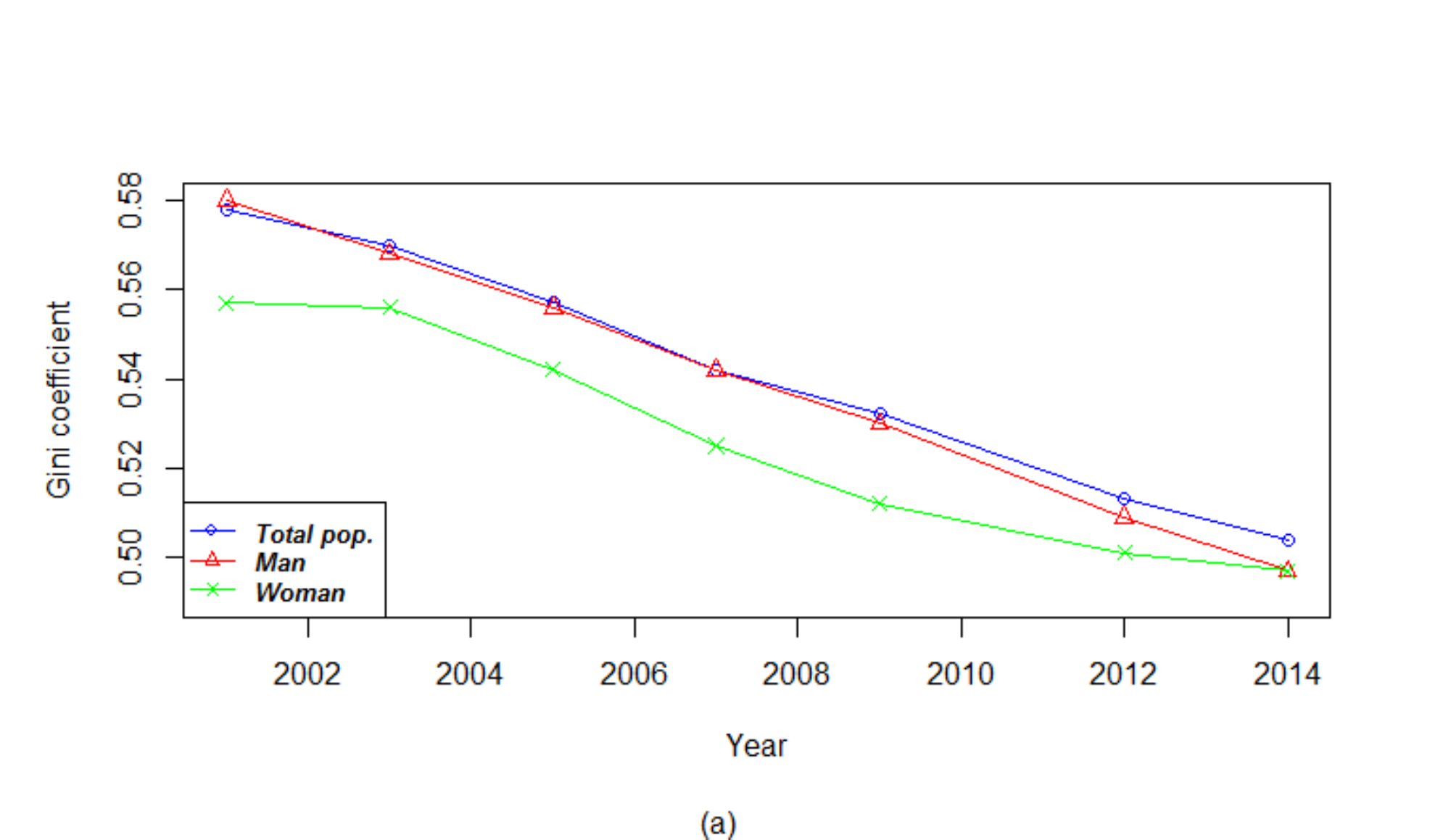}%
\end{minipage}\smallskip{}

\noindent\begin{minipage}[c]{1\columnwidth}%
\includegraphics[width=100mm]{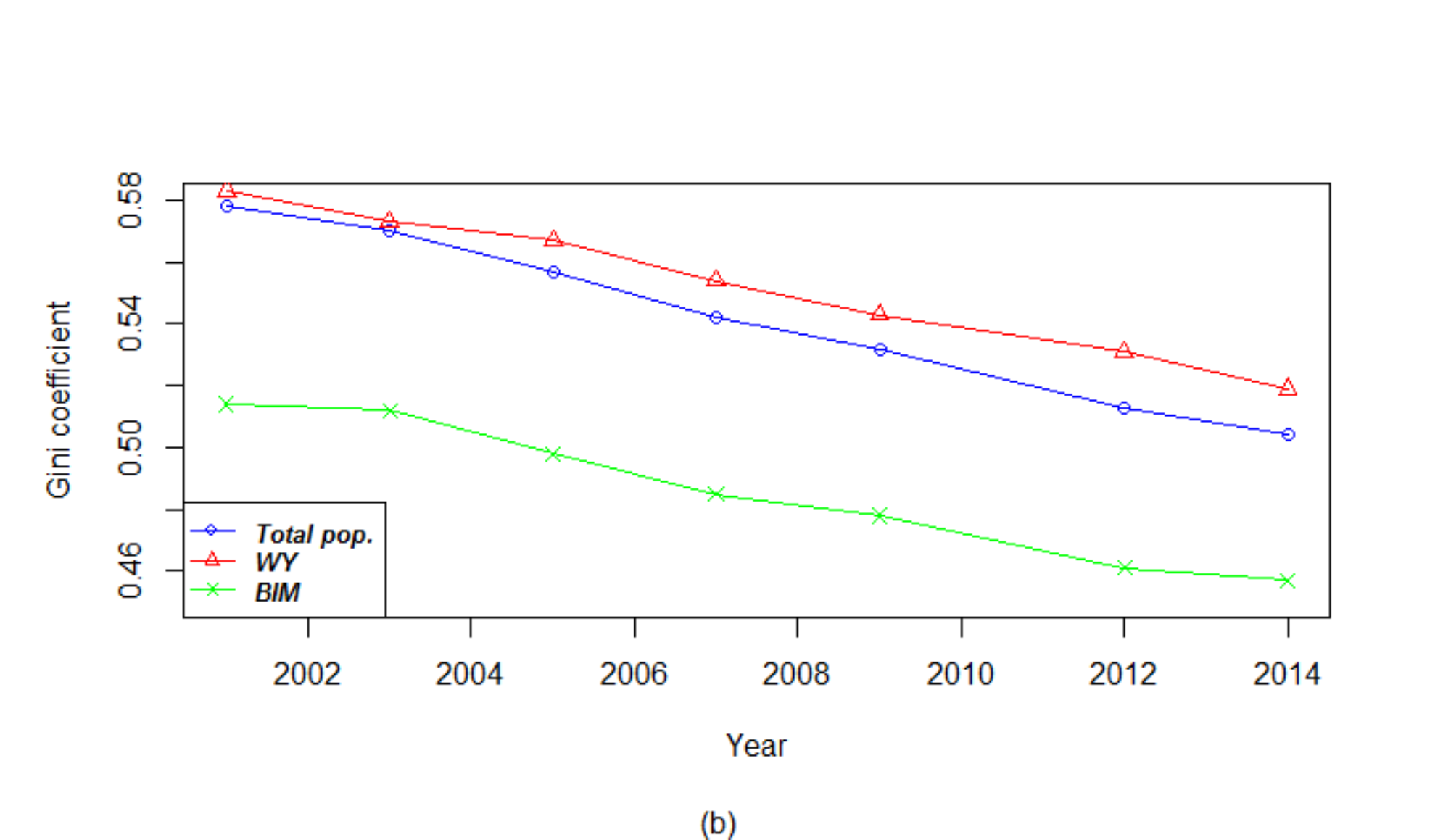}%
\end{minipage}

\noindent\begin{minipage}[c]{1\columnwidth}%
{\footnotesize{}Figure 5: (a) Gini coefficient for total population
divided by gender. (b) Gini coefficient for total population divided
by color, where BIM means black, indigenous, mixed race and WY means
white, yellow.}%
\end{minipage}
\end{figure}

For all the subgroups considered a significant improvement of the
Gini coefficient in these 14 years is observed. In specific for the
total population this inequality parameter changed from 0.578 to 0.504
an development of approximately 13\%. These findings suggest an decreasing
of Brazilian income inequality. One of the possible reasons for this
improvement is an increase in the number of people in the middle income
class where the value of the Gini coefficient is more sensitive to
changes. Moreover, as can be shown in Figure 5, historically-disadvantaged
subgroups (woman and BIM) have a better Gini coefficient. In the gender
case the difference between man's and woman's Gini coefficient over
the years selected was decreasing and in 2014 it became zero. Therefore,
another pertinent question has been raised, how is money distributed
among these subgroups? To answer this question, we evaluate the percentage
difference of the average income and the crossover income for each
subgroup in comparison with the total population. 

\begin{figure}[h!]
Percentage difference relative to the total population

\noindent\begin{minipage}[c]{1\columnwidth}%
\includegraphics[width=100mm]{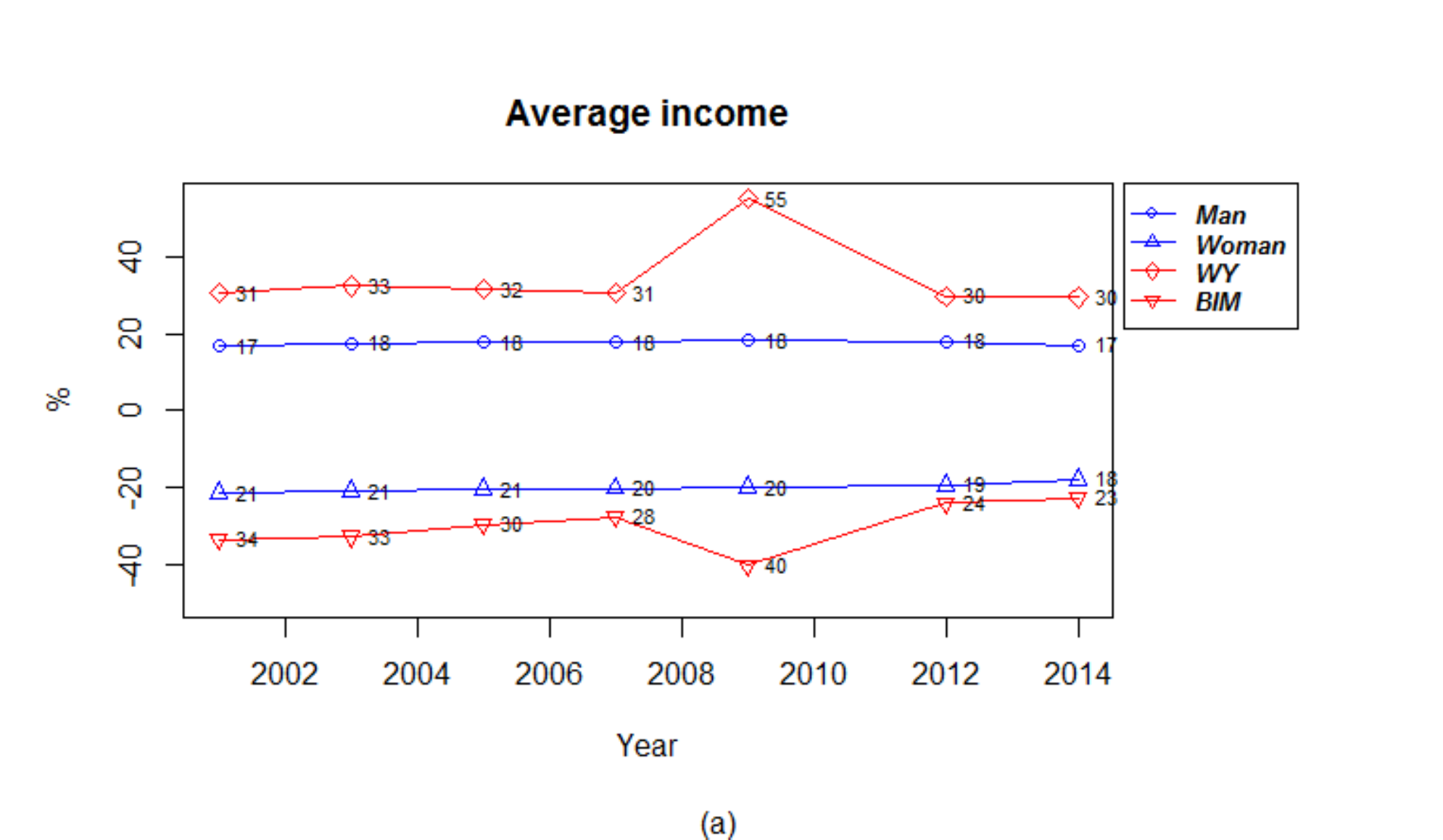}%
\end{minipage}\smallskip{}

\noindent\begin{minipage}[t][1\totalheight][c]{1\columnwidth}%
\includegraphics[width=100mm]{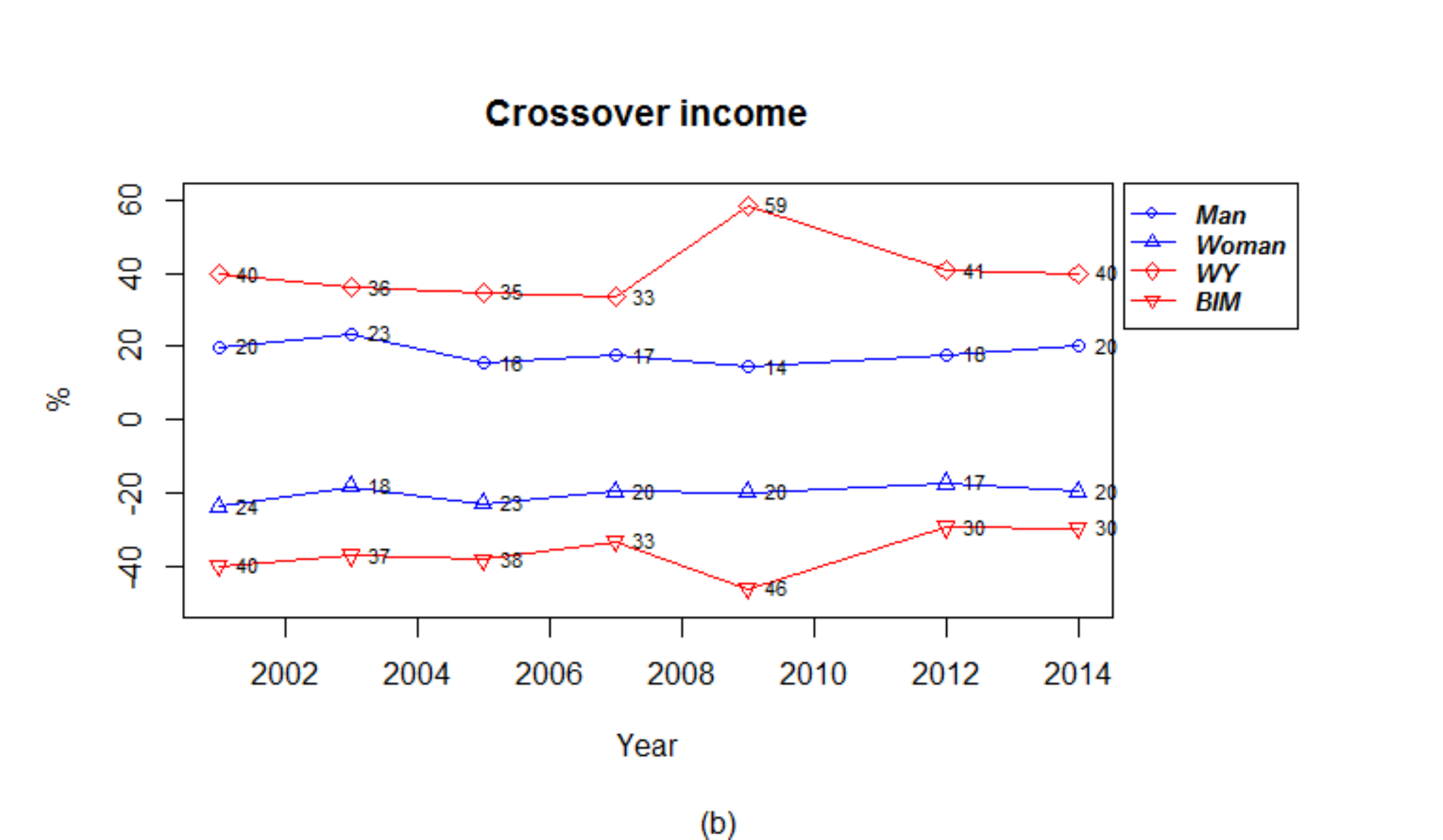}%
\end{minipage}

\noindent\begin{minipage}[t]{1\columnwidth}%
{\footnotesize{}Figure 6: (a) Average income percentage difference
from the total population. The standard error adopted for the graph
is 2\% and it was obtained by calculating the propagated error for
each individual measure where we chose the highest uncertainty among
all errors. (b) crossover income percentage difference relative to
the total population. The standard error is also about 2\% due to
ranging of 4\% to 6\% in high-income region. }%
\end{minipage}
\end{figure}

In Figure 6 it is indicated that the income inequality is more expressive
when the population is divided by color and also that the percentage
difference are larger for the crossover income than the average income.
Although their similar behavior, average income and crossover income
have no strong correlation. Focusing the gender dichotomy (blue lines),
there is no substantial shift towards a better income equality for
studied parameters if we take into account the standard errors. In
other words, despite the reduction over time of the Gini coefficient,
the fact that Brazilian women earn on average 40\% less than men remained
unchanged over the investigated years. When the Brazilian people is
separated by color/race (red lines)0, the percentage differences for
average income and crossover income with respects to the total population
are higher. Whites and yellows earn about 65\% to 50\% more than blacks,
mixed race and indigenous. Nevertheless, for BIM subgroup we could
notice a decrease of 10\% of its percentage difference in comparison
with the total population for average income and crossover income
over the selected years.

Another Gini coefficient limitation is that it does not contain information
about absolute national or personal incomes\cite{wikigini}. Hence,
even with a better Gini coefficient for historically-disadvantaged
subsets through Figure 6 we can see that these disadvantaged subgroups
of the Brazilian population have a much smaller monthly income than
the others. For this reason, it is possible to observe that the social
structure of men earning more than women and WY earning more than
BIM remained over the times series.

\begin{figure}[h!]
Time series of Pareto index

\noindent\begin{minipage}[t][1\totalheight][c]{1\columnwidth}%
\includegraphics[width=100mm]{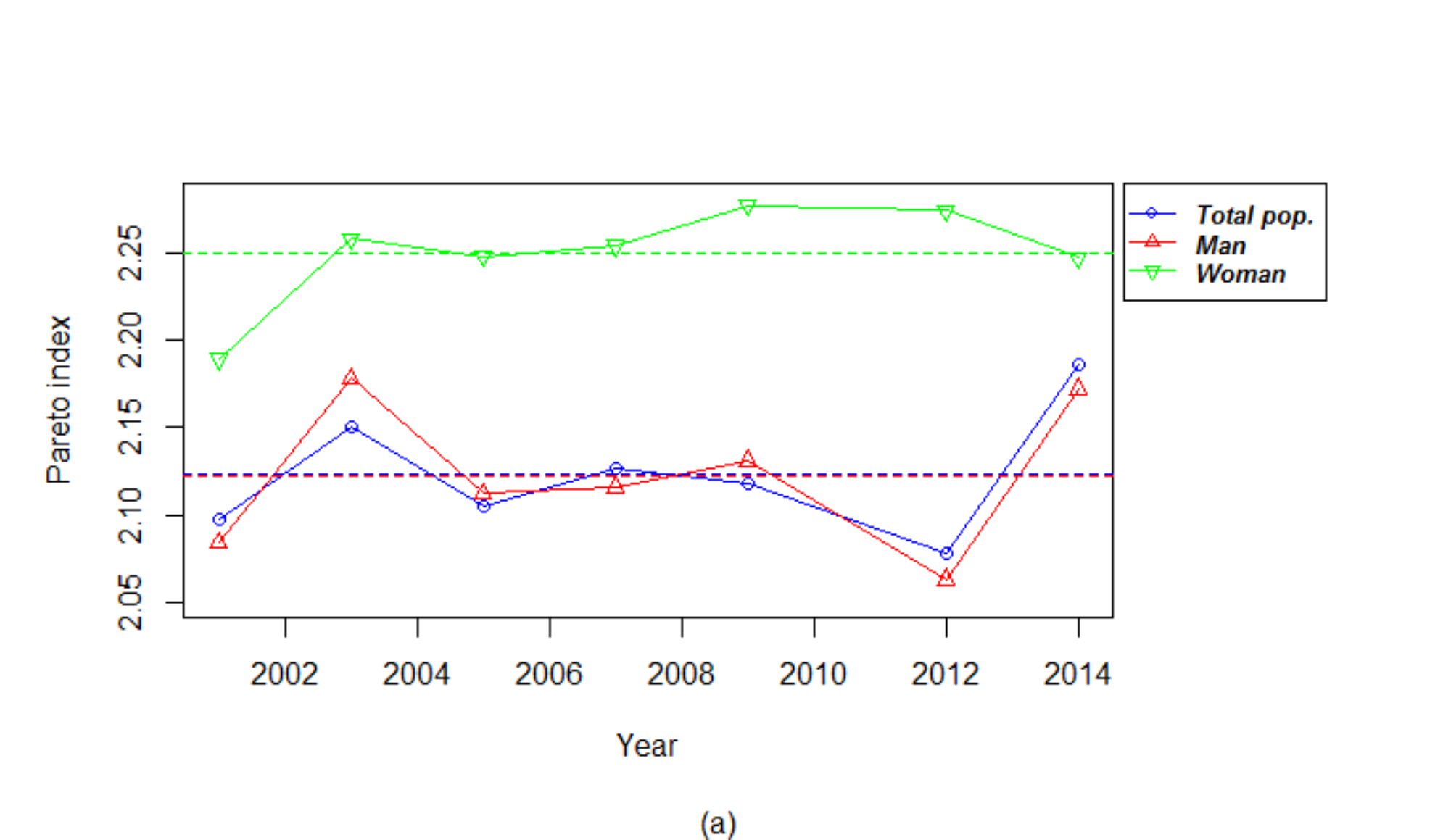}%
\end{minipage}

\smallskip{}

\noindent\begin{minipage}[t]{1\columnwidth}%
{\footnotesize{}\includegraphics[width=100mm]{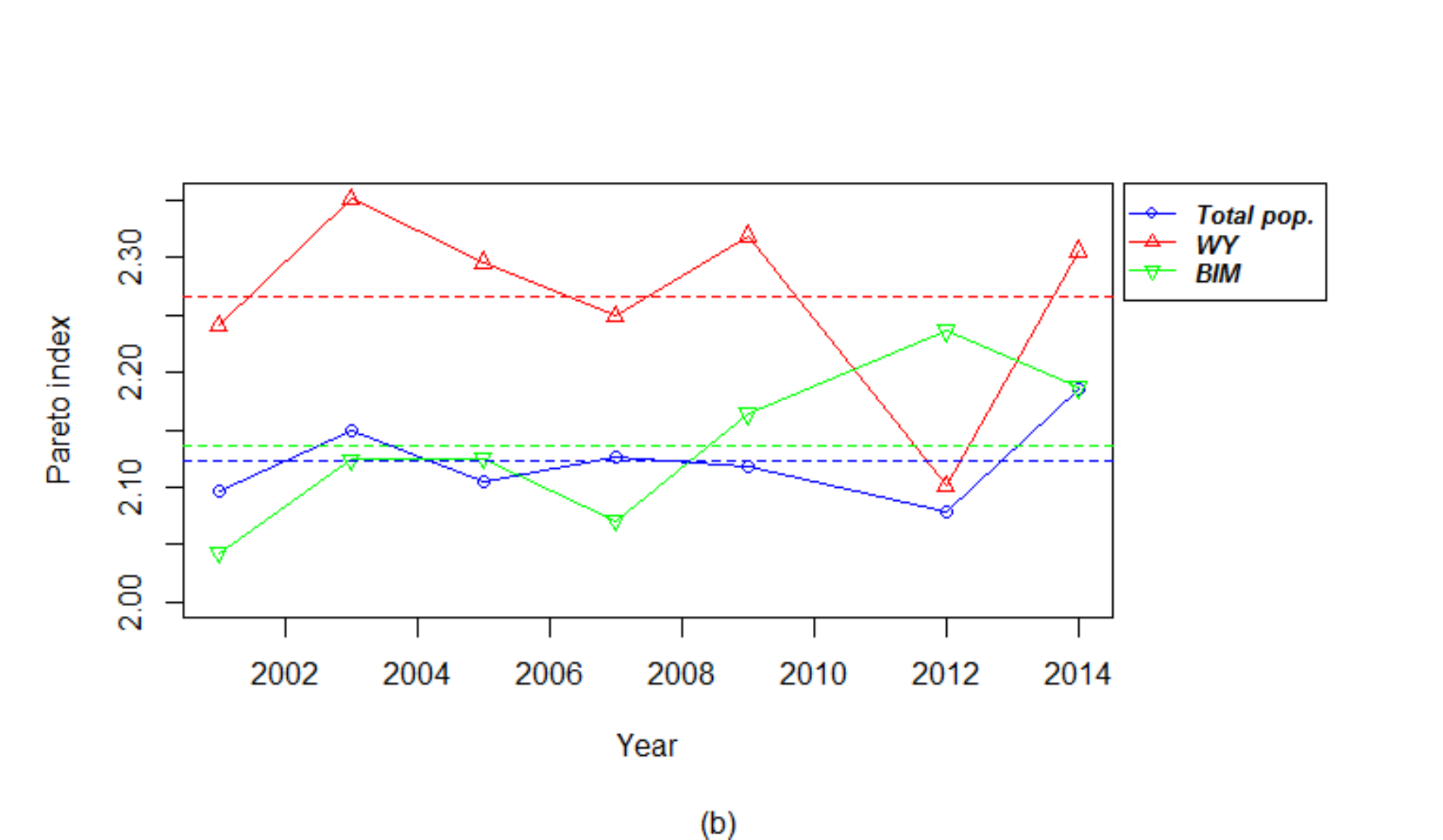}}%
\end{minipage}

\noindent\begin{minipage}[t]{1\columnwidth}%
{\footnotesize{}Figure 7: (a) Pareto index of total population divided
by gender. (b) Pareto index for total population divided by color.
The horizontals dashed lines represents the means which Pareto coefficients
of each subgroups fluctuates around.}%
\end{minipage}
\end{figure}

We use the Pareto exponent to investigated the income distribution
tail comprising roughly 5\% of the population that earn a high monthly
income. The insert of Figure 7 shows that for almost all subsets of
total population the Pareto index fluctuates around its respectively
means. This means are indicated for horizontal dashed lines. For the
group divided by sex and for the WY colors the slope coefficient has
magnitude $10^{-3}$. Otherwise, a slight increase is observed in
the BIM group where the slope coefficient is $10^{-2}$.

Thus, for the gender group thought unpaired T test with Welch's correction
the mean of Pareto coefficient over the selected years is significantly
higher for women than for men. This represents a better distribution
of income for the richer women. For the color group the situation
was inverted: the Pareto index for WY is statistically higher.


\begin{thebibliography}{10}

\bibitem{Clementi_2005}
F.~Clementi and M.~Gallegati.
\newblock Pareto's law of income distribution: Evidence for germany, the united
  kingdom, and the united states.
\newblock In {\em New Economic Windows}, pages 3--14. Springer Nature, 2005.

\bibitem{cg2005}
F.~Clementi and M.~Galleti.
\newblock Power law tails in the italian personal income distribution.
\newblock {\em Physica A: Statistical Mechanics and its Applications},
  350:427--438, 2005.

\bibitem{Dr_gulescu_2001}
A.~Dr{\u{a}}gulescu and V.~M. Yakovenko.
\newblock Exponential and power-law probability distributions of wealth and
  income in the united kingdom and the united states.
\newblock {\em Physica A: Statistical Mechanics and its Applications},
  299(1-2):213--221, oct 2001.

\bibitem{F2008}
M.~S. Faria, N.~S. Branco, and M.~H.~R. Tragtenberg.
\newblock Nonuniversal behavior for aperiodic interactions within a mean-field
  approximation.
\newblock {\em Physical Review E}, 77(4), apr 2008.

\bibitem{hd1972}
R.~Hoffmann and J.~Duarte.
\newblock The distribution of income in brazil.
\newblock {\em Rev. adm. empres}, 12:46--66, 1972.

\bibitem{De_Maio_2007}
F.~G.~D. Maio.
\newblock Income inequality measures.
\newblock {\em Journal of Epidemiology Community Health}, 61(10):849--852, oct
  2007.

\bibitem{ms1997}
L.~Moshe and S.~Solomon.
\newblock New evidence for the power-law distribution of wealth.
\newblock {\em Physica A: Statistical Mechanics and its Applications},
  242:90--94, 1997.

\bibitem{ibge}
B.~I. of~Geography and S.~. IBGE.
\newblock http://www.ibge.gov.br/home/.

\bibitem{S2014}
A.~Shaikh, N.~Papanikolaou, and N.~Wiener.
\newblock Race, gender and the econophysics of income distribution in the
  {USA}.
\newblock {\em Physica A: Statistical Mechanics and its Applications},
  415:54--60, dec 2014.

\bibitem{Silva_2005}
A.~C. Silva and V.~M. Yakovenko.
\newblock Temporal evolution of the
  {\textquotedblleft}thermal{\textquotedblright} and
  {\textquotedblleft}superthermal{\textquotedblright} income classes in the
  {USA} during 1983{\textendash}2001.
\newblock {\em Europhysics Letters ({EPL})}, 69(2):304--310, jan 2005.

\bibitem{s2006}
S.~Sinha.
\newblock Evidence for power-law tail of the wealth distribution in india.
\newblock {\em Physica A: Statistical Mechanics and its Applications},
  359:555--52, 2006.

\bibitem{s2002}
W.~Souma.
\newblock Physics of personal income.
\newblock 2002.
\newblock cond-mat/0202388v1.

\bibitem{wikigini}
Wikipedia.
\newblock https://goo.gl/WxohyG.

\bibitem{w2010}
M.~Wrzeszcz, M.~Slomski, and S.~Rybarczyk.
\newblock Measures of inequality and empirical results.
\newblock https://goo.gl/6OJDD0, 2010.

\end{thebibliography}
\end{document}